\documentclass[12pt,a4paper]{article}

\usepackage{amsmath,amssymb}
\usepackage{graphicx}
\usepackage[nosort]{cite}
\usepackage[hyperref,parsep]{collect}

\makeatletter
\let\old@startsection=\@startsection
\renewcommand{\@startsection}[6]{\old@startsection{#1}{#2}{#3}{#4}{#5}{#6\mathversion{bold}}}
\makeatother

\makeatletter
\let\old@makecaption=\@makecaption
\def\@makecaption{\small\old@makecaption}
\makeatother

\let\oldPhi=\Phi
\let\oldPsi=\Psi
\let\oldGamma=\Gamma
\let\oldDelta=\Delta
\let\oldSigma=\Sigma
\let\oldTheta=\Theta
\let\oldPi=\Pi
\renewcommand{\Phi}{\mathnormal{\oldPhi}}
\renewcommand{\Psi}{\mathnormal{\oldPsi}}
\renewcommand{\Gamma}{\mathnormal{\oldGamma}}
\renewcommand{\Sigma}{\mathnormal{\oldSigma}}
\renewcommand{\Delta}{\mathnormal{\oldDelta}}
\renewcommand{\Theta}{\mathnormal{\oldTheta}}
\renewcommand{\Pi}{\mathnormal{\oldPi}}

\newcommand{\sign}{\mathop{\mathrm{sign}}}
\renewcommand{\Re}{\mathop{\mathrm{Re}}}

\newcommand{\order}[1]{\mathcal{O}(#1)}

\newcommand{\contour}{\mathcal{C}}
\newcommand{\Integers}{\mathbb{Z}}

\ifx\genfrac\sdflkaj
\newcommand{\atopfrac}[2]{{{#1}\above0pt{#2}}}
\else
\newcommand{\atopfrac}[2]{\genfrac{}{}{0pt}{}{#1}{#2}}
\fi
\newcommand{\sfrac}[2]{{\textstyle\frac{#1}{#2}}}
\newcommand{\half}{\sfrac{1}{2}}
\newcommand{\ihalf}{\sfrac{i}{2}}
\newcommand{\quarter}{\sfrac{1}{4}}

\newcommand{\indup}[1]{_{\mathrm{#1}}}

\newcommand{\alg}[1]{\mathfrak{#1}}

\newcommand{\lrbrk}[1]{\left(#1\right)}
\newcommand{\bigbrk}[1]{\bigl(#1\bigr)}
\newcommand{\Bigbrk}[1]{\Bigl(#1\Bigr)}

\newcommand{\set}[1]{\{#1\}}

\newcommand{\pint}{\makebox[0pt][l]{\hspace{3.4pt}$-$}\int}

\newcommand{\sheetsl}[1][]{\makebox[0pt][l]{\hspace{0.06em}$/$}#1p}

\newcommand{\nln}{\nonumber\\}

\newcommand{\earel}[1]{\mathrel{}&\hspace{-2\arraycolsep}#1\hspace{-2\arraycolsep}&\mathrel{}}
\newcommand{\eq}{\earel{=}}

\def\[{\begin{equation}}
\def\]{\end{equation}}
\def\<{\begin{eqnarray}}
\def\>{\end{eqnarray}}

\makeatletter
\def\mr@ignsp#1 {\ifx\:#1\@empty\else #1\expandafter\mr@ignsp\fi}%
\newcommand{\multiref}[1]{\begingroup%\let\protect\string%
\xdef\mr@no@sparg{\expandafter\mr@ignsp#1 \: }%
\def\mr@comma{}%
\@for\mr@refs:=\mr@no@sparg\do{\mr@comma\def\mr@comma{,}\ref{\mr@refs}}%
\endgroup}
\makeatother

\newcommand{\hypref}[2]{\ifx\href\asklfhas #2\else\href{#1}{#2}\fi}

\newcommand{\figref}[1]{Fig.~\multiref{#1}}
\renewcommand{\eqref}[1]{(\multiref{#1})}

\ifx\href\asklfhas\newcommand{\href}[2]{#2}\fi

\begin{document}
\begin{flushright}\footnotesize
\texttt{hep-th/0506243}\\
\texttt{NORDITA-2005-44}\\
\texttt{PUTP-2164}
\end{flushright}
\vspace{0cm}

\begin{center}
{\Large\textbf{\mathversion{bold}
Fluctuations and Energy Shifts\\
in the Bethe Ansatz}\par}
\vspace{1cm}

\textsc{N.~Beisert$^a$ and L.~Freyhult$^b$}
\vspace{5mm}

\textit{$^a$ Joseph Henry Laboratories\\
Princeton University\\
Princeton, NJ 08544, USA}\vspace{3mm}

\textit{$^b$ NORDITA\\
Blegdamsvej 17\\
DK-2100 Copenhagen \O, Denmark}\vspace{3mm}
\vspace{3mm}

\texttt{nbeisert@princeton.edu}\\
\texttt{freyhult@nordita.dk}\par\vspace{1cm}

\textbf{Abstract}\vspace{7mm}

\begin{minipage}{12.7cm}
We study fluctuations and finite size corrections 
for the ferromagnetic thermodynamic limit in the 
Bethe ansatz for the Heisenberg XXX$_{1/2}$ spin chain,
which is the AdS/CFT dual of semiclassical spinning strings.
For this system we derive the standard quantum mechanical formula 
which expresses the energy shift as a sum over fluctuation energies. 
As an example we apply our results to the simplest, 
one-cut solution of this system and derive its spectrum of fluctuations.
\end{minipage}

\end{center}
\vspace{1cm}
\hrule height 0.75pt
\vspace{1cm}

The problem of finding planar anomalous dimensions 
of local operators in large-$N$ maximally supersymmetric gauge theory 
is simplified drastically by its apparent integrability \cite{Minahan:2002ve,Beisert:2003tq}
(see \cite{Beisert:2004ry} for a review).
Using suitable Bethe ans\"atze has drastically simplified the comparison 
to plane wave strings \cite{Callan:2003xr}
(see \cite{Swanson:2005wz} for a review)
and classical spinning strings \cite{Gubser:2002tv,Frolov:2002av}
(see \cite{Tseytlin:2003ii,Beisert:2004yq,Tseytlin:2004xa,Zarembo:2004hp} for reviews)
in the context of the AdS/CFT conjecture.
One interesting problem in this context is the first order correction 
to classical spinning string energies.
In string theory, this is a one-loop quantum correction in
the string sigma model. 
In gauge theory it corresponds to finite-size effects 
in a thermodynamic limit of long local operators.
At this order, 
there are two types of effects that can be considered: 
The spectrum of fluctuation energies around some
classical solution has been investigated 
in \cite{Frolov:2003tu,Beisert:2003xu,Freyhult:2004iq}.
The other effect is the energy shift
of the classical solution, 
c.f.~\cite{Frolov:2003tu,Lubcke:2004dg,Frolov:2004bh,Park:2005ji,Freyhult:2005fn,Beisert:2005mq,Hernandez:2005nf}.
In fact, in string theory these two effects are related by the
general quantum mechanical formula for the first-order energy shift
\[\label{eq:Shift}
\delta E=\frac{1}{2}\sum\nolimits_k e_k.
\]
Here $e_k$ is the energy (shift) of a fluctuation mode labelled by $k$ 
around some classical state 
and $\delta E$ is the one-loop energy shift of the same state.
Conversely, in gauge theory, the computations of $e_k$ and $\delta E$ in
\cite{Beisert:2003xu,Freyhult:2004iq} and
\cite{Beisert:2005mq,Hernandez:2005nf} are not 
related in an obvious way.
The purpose of this note is to derive a formula similar
to \eqref{eq:Shift} in gauge theory.
This might facilitate the computation of $\delta E$ 
or, for the sake of comparison, even make it obsolete.
Here we focus on the case of the $\alg{su}(2)$ sector
of two charged scalar fields which is dual to
the Heisenberg XXX$_{1/2}$ quantum spin chain
\cite{Minahan:2002ve}.
\bigskip

The Bethe equation for the Heisenberg XXX$_{1/2}$ spin chain in logarithmic form
reads
\[\label{eq:BetheExact}
2\pi n_k=-i\sum_{\textstyle\atopfrac{j=1}{j\neq k}}^{K}\log\frac{u_k-u_j+i}{u_k-u_j-i}
-iL\log\frac{u_k-\sfrac{i}{2}}{u_k+\sfrac{i}{2}}\,.
\]
In addition to the Bethe equation, gauge theory states must obey the
cyclicity constraint
\[\label{eq:Cyclicity}
2\pi m=-i\sum_{k=1}^{K}\log\frac{u_k+\ihalf}{u_k-\ihalf}\,.
\]
The energy and anomalous dimension is given by the formula
\[
E=\sum_{k=1}^K \frac{1}{u_k^2+\quarter}\,,\qquad D=L+\frac{\lambda\,E}{8\pi^2}+\ldots\,.
\]
Here we consider the (ferromagnetic low-energy) thermodynamic limit $L\to\infty$.
In this limit the Bethe roots scale as $u\sim L$ and 
condense on curves $\contour_a$ in the complex plane 
with density $\rho(u)\sim 1$ \cite{Sutherland:1995aa,Beisert:2003xu}.
The expansion of the Bethe equation at $\order{1/L}$ reads
\cite{Beisert:2005mq,Hernandez:2005nf}
\<\label{eq:Bethe1L}
2\pi n_a\eq
\pint_{\contour} \frac{2\,du'\,\rho(u')}{u-u'}
-\frac{L}{u}
+\sum^{+\infty}_{k=-\infty}
\frac{\rho'(u)}{\rho(u)+ik}
+\ldots
\nln\eq
\pint_{\contour} \frac{2\,du'\,\rho(u')}{u-u'}
-\frac{L}{u}
+\pi\rho'(u)
\coth(\pi \rho(u))
+\ldots\qquad
\mbox{for }u\in \contour_a.
\>
Note that the first two
terms represent the long-range ($u_k-u_j\sim L$) 
contributions from \eqref{eq:BetheExact}.
The latter term is a short-range ($u_k-u_j\sim 1$)
contribution and it can be considered as an anomaly.%
\footnote{We are grateful to V.~Kazakov for discussions on this point;
see also \cite{Beisert:2005di}.}

In this form the derivation of concrete $\order{1/L}$
energy shifts appears somewhat tedious.
For example, the equation makes explicit reference 
to the shape of the contours $\contour_a$ which is determined
by the condition that the density $du\,\rho(u)$ is real and positive.
When the curve extends into the complex plane, it does not have
a simple shape.%
\footnote{In practice one analytically continues some parameter such that
the curve flips onto the real axis.}
To avoid the problem of an unknown curve, it is convenient 
to introduce a (leading-order) quasi-momentum $p_0(u)$
\cite{Kazakov:2004qf}
\[\label{eq:quasi}
p_0(u)=
\int_{\contour} \frac{du'\,\rho(u')}{u'-u}
+\frac{L}{2u}\,.
\]
The Bethe equation is then interpreted as an 
integrality condition on cycles of the curve $dp$, 
independent on the choice of branch cuts in $p$.
Sticking to the above form, the Bethe equation now reads
\[\label{eq:BetheSemi}
0=\sheetsl_0(u)+\pi n_a-\half\pi\rho'(u)
\coth(\pi \rho(u))
+\ldots\qquad\mbox{for }u\in \contour_a,
\]
where $\sheetsl(u)=\half p(u+i\epsilon)+\half p(u-i\epsilon)$ is the
principal value of $p(u)$.
This equation still makes reference to the density $\rho(u)$.
We can eliminate it as the discontinuity of $p_0$ across $\contour_a$ in
\eqref{eq:quasi}
\<
\rho(u)\eq\bigbrk{p_0(u+i\epsilon)-p_0(u-i\epsilon)}/2\pi i
\nln\eq
\bigbrk{p_0(u+i\epsilon)-p_0(u-i\epsilon)\pm 2\sheetsl_0(u)\pm2\pi n_a+\ldots}/2\pi i
\nln\eq
\mp ip_0(u\pm i\epsilon)/\pi\mp in_a+\ldots\,.
\>
In the second line we have substituted the Bethe equation \eqref{eq:BetheSemi}
and dropped subleading terms. We substitute back into \eqref{eq:BetheSemi}
and find
\[
2\pi n_a=
-2\sheetsl_0(u)\,
+p'_0(u\pm i\epsilon)\cot(p_0(u\pm i\epsilon))
+\ldots\qquad\mbox{for }u\in \contour_a.
\]
Interestingly, the reference to the mode number $n_a$ has dropped out and
the sign ambiguity merely tells us to evaluate the second term 
slightly away from $u$ on either side of the curve $\contour_a$.
We can therefore introduce the full quasi-momentum
\[
p(u)=
p_0(u)\,
-\half p'_0(u)\cot(p_0(u))+\ldots
\]
which obeys $\sheetsl(u)=-\pi n_a$ for $u\in\contour_a$. 

The extra term is now expressed through $p_0(u)$ 
and therefore has an analytic continuation 
almost everywhere in the complex plane. 
We can alternatively represent such a function through its singularities
which we shall now investigate. The cotangent has poles at $\pi \Integers$, 
let us therefore denote the solutions to $p_0(u_k)=-\pi k$ by $u_k$.
The residue at $u_k$ is $-\half$. 
Furthermore, we should consider the branch points $u_a^{\pm}$ of $p_0(u)$,
i.e.~the pair of endpoints of the contours $\contour_a$. 
At these points we have 
$p_0(u)=-\pi n_a+\ast \sqrt{u-u^\pm_a}+\ldots$
and therefore $p_0(u^\pm_a)=-\pi n_a$ is satisfied as well.
Due to the square root the residues are $-\quarter$.
These are all the singularities, only the constant part remains to be fixed. 
At $u=\infty$ we have $p_0(u)\sim 1/u$ and
therefore also the extra term vanishes.
Assembling this information we obtain the relation
\[
p(u)=p_0(u)+
\sum_{k\not\in\set{n_a}}\lrbrk{\frac{1/2}{u_k-u}+\frac{1/2}{u}}
+\sum_{a,\pm}\lrbrk{\frac{1/4}{u^\pm_a-u}+\frac{1/4}{u}}+\ldots\,.
\]
Note that the first term can also be interpreted as
the contribution from all the unoccupied cuts with mode number 
$k\not\in\set{n_a}$. As their length is zero, 
the would-be branch points degenerate
$u^+_k=u^-_k=u_k$ and each of the poles at $u^\pm_k$ 
contributes half the residue of $u_k$.
\medskip

We can now relate this expression to the spectrum
of fluctuations. A fluctuation is obtained by adding 
a single Bethe root to the original distribution of roots \cite{Beisert:2003xu}.
This leads to a energy shift of order $1/L^2$.
There are two principal contributions: 
Firstly, the new root contributes directly to the energy.
Excitingly, the above $u_k$ is precisely the leading-order position 
of an additional Bethe root with mode number $k$.
Secondly, the new potential due to the new root distorts 
the original distribution of roots. This is described by a 
deformed quasi-momentum 
\[\label{eq:quasideform}
p(u)=
p_0(u)\,
+\frac{1}{u_k-u}+\ldots\,.
\]
So this is very suggestive for the applicability of the 
quantum mechanical formula \eqref{eq:Shift}.
There are however a few modifications we have to take 
into account. 
Most importantly, we do not consider the direct contribution to the energy. 
Furthermore, we add a pole at $u=0$ for each mode.
As $u_k\sim 1/k$ for $k\to \infty$ this can be interpreted
as subtracting the contribution for $k=\infty$.
The accurate formula for the energy shift is therefore
\[\label{eq:ShiftReg}
\delta E=\frac{1}{2}\sum_{k=-\infty}^{+\infty} 
\lrbrk{\tilde e_k-\tilde e_\infty},
\qquad
\tilde e_k=e_k-\frac{1}{2(u_k^+)^2}-\frac{1}{2(u_k^-)^2}\,.
\]
The additional terms can be understood as a regularisation 
of \eqref{eq:Shift}:
For large mode numbers $k$, the fluctuation energy
asymptotes as $e_k=k^2+a k+b+c/k+\order{1/k^2}$.
Clearly \eqref{eq:Shift} diverges in this case.
However, the Bethe root asymptotes as $u_k^\pm=1/k+\order{1/k^2}$
and precisely cancels the leading $k^2+ak$
in $\tilde e_k=\tilde b+\tilde c/k+\order{1/k^2}$. 
Furthermore $\tilde e_\infty=\tilde b$, so the summand is $\order{1/k}$.
Finally, we need to recall that the sum in 
\eqref{eq:Bethe1L} implies symmetric boundaries;
the sum therefore converges.
\bigskip

\begin{figure}\centering
\includegraphics{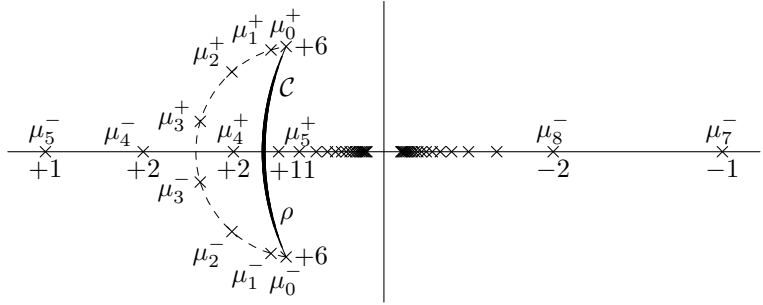}
\caption{The one-cut solution $\contour,\rho$ with $n=6$ 
for some fixed value of $\alpha=K/L$
and the distribution of fluctuation roots $\mu_k^\pm$.
The mode number of $\mu^+_k$ is $6+k$ for $k> k\indup{c}=4$
for all other $\mu^+_k,\mu^-_k$ it is $6-k$.
Some of these mode numbers are indicated in the figure.
As compared to the vacuum with $\alpha=0$, 
the roots between $\mu_4^-$ and $\mu_4^+$, as well as their mode numbers, 
are distorted strongly by the cut.}
\label{fig:Roots}
\end{figure}

As an example we consider adding fluctuations to the one-cut solution 
discussed in \cite{Kazakov:2004qf,Lubcke:2004dg,Beisert:2005mq}. 
The mode number of the cut is $n$ which we assume to be positive 
without loss of generality.
Keeping the first order terms in the $1/L$ expansion 
the Bethe equation for the one-cut solution in the thermodynamic limit is
\[
\frac{1}{q}+2\pi n+2\pint_{\contour_0}\frac{dq'\,\rho_0(q')}{q'-q}=0,
\]
here we have introduced the rescaled roots $u=Lq$.
The resolvent is then 
\[\label{eq:resolvent}
G_0(q)=\int_{\contour_0}\frac{dq'\,\rho_0(q')}{q'-q}
=\frac{1}{2q}\left(-1-2\pi nq+\sqrt{(1+2\pi nq)^2-8\pi n\alpha q}\right)
\]
where $\alpha=K/L$. The cut $\contour_0$ is on the left side of the complex
plane, vertical and curved away from the origin, see \figref{fig:Roots}.
The energy (and anomalous dimension) of this solution is
\[
E_0=\frac{4\pi^2 n^2 \alpha(1-\alpha)}{L}\,,\qquad
\delta D=\frac{\lambda\,E}{8\pi^2}=\frac{\lambda n^2 \alpha(1-\alpha)}{2L}+\ldots\,.
\]
Furthermore, the cyclicity constraint \eqref{eq:Cyclicity} requires
$2\pi m=G_0(0)$, whereas \eqref{eq:resolvent} gives 
$G_0(0)=2\pi n \alpha$, i.e.~we have the constraint $\alpha=-m/n$.

We introduce a fluctuation by placing a single root at position $\mu$.
We assign the mode number $n+s k$ with $k\geq 0$ and $s=\pm 1$; 
the position up to $1/L$ corrections is determined by
\[\label{eq:muposeq}
\frac{1}{\mu}+2\pi (n+sk)+2G_0(\mu)=0.
\]
Using \eqref{eq:resolvent} we find 
\[\label{eq:mupos}
\frac{1}{\mu}=-2\pi\left((1-2\alpha)n\pm\sqrt{k^2-4n^2\alpha(1-\alpha)}\right).
\]
Note that the sign of $\mu$ in \eqref{eq:mupos} 
is not necessarily the same as $s$ in \eqref{eq:muposeq}. 
Let us therefore investigate the sign $s^\pm_k=s$ 
for the mode number of the Bethe root $\mu^\pm_k=\mu$  
as we increase $\alpha$, c.f.~\figref{fig:Roots}.
At $\alpha=0$ we have $G_0(q)=0$ and therefore 
$\mu_k^\pm=-1/2\pi (n\pm k)$, 
i.e.~$s^\pm_k=\pm 1$.
When we increase $\alpha$, 
the branch points $\mu^\pm_0$ of $\contour$ will split
up into the complex plane.
Furthermore, all the fluctuations $\mu_k^\pm$ will be attracted towards
the curve $\contour$.
At some value of $\alpha$, 
the fluctuation $\mu_{1}^{+}$ 
will cross the curve $\contour$;%
\footnote{Here, the density of
roots near the real axis is $\rho=i$. 
It is conceivable that this signals some
kind of instability or phase transition of the classical solution 
due to the pole in the interaction kernel.
Possibly, the solution forms a condensate of roots with 
constant density $\rho=i$, c.f.~\cite{Beisert:2005mq}.
In this case our (as well as the classical) analysis does not apply when $k\indup{c}>0$. 
To proceed, and in agreement with the treatment on the string side, 
we shall assume the classical one-cut solution to remain valid when $k\indup{c}>0$.}
it turns out that the crossing
is always from the concave to the
convex side of $\contour$.
During the crossing the mode number 
of $\mu_{1}^{+}$ changes from $n+1$ to $n-1$, the same as for $\mu_{1}^{-}$.
Somewhat later both roots $\mu_{1}^{\pm}$ will collide and 
branch off into the complex plane as a conjugate pair. 
As we increase $\alpha$ further, 
more and more roots $\mu^{+}_{k}$ will cross $\contour$ 
and later move into the complex plane together with
$\mu^{-}_{k}$. 
The sign for the mode number $n+s^\pm_k k$
of $\mu^{\pm}_{k}$ is therefore
$s^+_k=\sign(k-k\indup{c}-\epsilon)$
where $k\indup{c}$ is the critical mode number
and $s^-_k=-1$.
Note that $k\indup{c}=[|\rho|]$ is given by the integer part
of the density where $\contour$ crosses the real axis.

The Bethe root $\mu$ will contribute to the fluctuation energy directly 
but it will also modify the density on the cut. 
This will lead to another contribution of order $1/L^2$ to the energy. 
In order to obtain this correction we need to solve the equation
\eqref{eq:quasideform}
\[\label{eq:ModBethe}
\frac{1}{q''}+2\pi n-\frac{2}{L}\,\frac{1}{q''-\mu}+2\pint_{\contour}\frac{\rho(q')\,dq'}{q'-q''}=0.
\]
Technically we do this using the method introduced in \cite{Lubcke:2004dg}.
We integrate the equation by $dq''\rho(q'')/(q-q'')$ 
to find an algebraic equation for the deformed resolvent $G(q)$
\[\label{eq:algebraic}
G(q)^2+\left(\frac{1}{q}+2\pi n-\frac{2}{L}\,\frac{1}{q-\mu}\right)G(q)
-\frac{G(0)}{q}
+\frac{2}{L}\,\frac{G(\mu)}{q-\mu}=0.
\]
Integrating \eqref{eq:ModBethe} over $dq''\,\rho(q'')$ instead, 
we obtain 
\[
G(0)=-2\pi n\alpha+\frac{2}{L}\,G(\mu).
\]
By expanding \eqref{eq:algebraic} 
and approximating $G(\mu)$ by $G_0(\mu)$ in \eqref{eq:muposeq}
we find
\[
G'(0)=(2\pi)^2n^2\alpha(1-\alpha)
+\frac{1}{L}
\lrbrk{
-\frac{1}{\mu^2}
-\frac{2\pi sk}{\mu}
+(2\pi)^2n(n+sk)(1-2\alpha)
}
\]
which contributes to the fluctuation energy,
\[\label{eq:flucteng}
e^\pm_{n+sk}=\frac{\delta G'(0)}{L}+\frac{1}{L^2\mu^2}
=
\frac{(2\pi)^2}{L^2}\left(n(n+2sk)(1-2\alpha)\pm sk \sqrt{k^2-4n^2\alpha(1-\alpha)}\right).
\]
Strictly speaking the above result is only valid 
when $k^2-4n^2\alpha(1-\alpha)>0$. 
In the Bethe ansatz the existence of a well defined quantum state requires 
complex Bethe roots to come in pairs with their complex conjugates. 
As $k^2-4n^2\alpha(1-\alpha)$ becomes negative 
the added root will move into the complex plane. 
The contribution to the energy from the conjugate pair, 
$\mu^+_{k}$ and $\mu^-_{k}$, will then be $e^+_{n+sk}+e^-_{n+sk}=2\Re e^\pm_{n+sk}$. 
A single root moving into the complex plane and causing an instability 
will not correspond to a well-defined quantum state in the spin chain picture, 
nevertheless this will correspond to the instability in the dual string result. 
Note that the term under the square root becomes negative 
when $\alpha\indup{c}\geq\frac{1}{2}(1-\sqrt{1-1/n^2})$ 
which is always true when the cyclicity constraint $\alpha=-m/n$ holds.

The formula \eqref{eq:flucteng} reproduces two known energies:
The fluctuation with mode number zero, i.e.~$\mu_n^-$, is in the same multiplet 
as the classical state; we thus have a vanishing energy shift $e_{n-n}^-=0$.
For mode number $n$ we have merely added a Bethe root to the existing cut.
Therefore $\half e^+_n+\half e^-_n=(2\pi n/L)^2(1-2\alpha)=(\partial E/\partial \alpha)/L=(\partial E/\partial K)|_L$.
Equation \eqref{eq:flucteng} also agrees with the 
analytically continued result from string theory \cite{Park:2005ji}:
Note that unlike in our computation, the fluctuation in \cite{Park:2005ji}
does not carry a $\alg{su}(2)$ charge. Instead of adding a Bethe root, we have to move
one from $k=0$ to the desired position, i.e.~we have to subtract $e_{n+0}$.
The remainder agrees with (C.1) in \cite{Park:2005ji}. To compare to 
(4.18) in \cite{Park:2005ji}, we have to recall that a single excitation is not
physical due to the cyclicity constraint \eqref{eq:Cyclicity}. 
In a composite physical state
all terms linear in $k$ cancel out.%
\footnote{We thank A.~Tseytlin for this explanation.}
\smallskip

The part relevant for computing the energy shift of the ground state is
\<\label{eq:etilde}
\tilde e^\pm_{n+sk}\eq e^\pm_{n+sk}-\frac{1}{L^2\mu^2}=\frac{\delta G'(0)}{L}
\\\nonumber\eq
\frac{(2\pi)^2}{L^2}\left(n^2\alpha(6-8\alpha)
+(sk-2n(1-2\alpha))\Bigbrk{\pm\sqrt{k^2-4n^2\alpha(1-\alpha)}-sk}\right).
\>
It follows from \eqref{eq:etilde} with $s^\pm_\infty=\pm1$ that 
\[
\tilde{e}_\infty=\frac{(2\pi)^2}{L^2}\,n^2\alpha(4-6\alpha)
\]
which equals
$\tilde{e}_\infty=-2\,\partial E/\partial L|_K$.
After symmetrization with respect to $k\mapsto -k$,
the $1/L$ correction to the energy is
\[
\delta E=
\frac{(2\pi)^2}{2L^2}\sum_{k=-\infty}^{\infty} 
\left(2n^2\alpha(1-\alpha)-k^2+|k|\cdot
  \begin{cases}-2n(1-2\alpha)&\mbox{for }|k|\leq k\indup{c},\\
  \sqrt{k^2-4n^2\alpha(1-\alpha)}&\mbox{for }|k|> k\indup{c}\end{cases}\right).
\]
This expression coincides almost with the result obtained in \cite{Beisert:2005mq} 
by direct computation of the energy shift of the classical solution.
The only difference is related to the instability and the 
subcritical modes $|k|\leq k\indup{c}$, an effect which cannot be
seen in the analytical continuation of \cite{Beisert:2005mq}.
\bigskip

In conclusion we have found that the energy shift 
obtained in \cite{Beisert:2005mq,Hernandez:2005nf} 
from considering finite size effects and the contribution 
from fluctuations around a classical solution 
in the gauge theory \cite{Beisert:2003xu,Freyhult:2004iq} can be related. 
This gives two alternative methods for computing $1/L$ corrections. 
It also shows that the spin chain in the thermodynamic limit can be considered a
quantum mechanical system where the standard formula for the first-order 
energy shift in $1/L$ applies. A more difficult, nevertheless very interesting project 
would be to extend the current result to second order in $1/L$.

Furthermore our result makes explicit the fact that it is enough 
to know the branch points of the cuts where the Bethe roots condense, 
not their explicit shape. 
This is expected as the dual description 
of the finite size contributions are obtained considering fluctuations 
around a classical string solution. We do not expect to need 
the full knowledge of the position of the cuts unless we examine 
the dual of a true quantum computation in string theory.
There is, however, a slight limitation to this point of view:
Close to the regime where instabilities occur 
we need to know the precise position of cuts in order to find 
the correct energy shifts. 
This point clearly requires further investigation.

It would be interesting to establish a similar relation 
between the energy shift and the fluctuations 
also in other sectors of the theory.
When we go to the complete model \cite{Beisert:2003yb,Beisert:2005di} and include fermions, 
the duality to string theory suggests that we can drop the 
regularization terms in \eqref{eq:ShiftReg} to 
obtain a simplified formula similar to \eqref{eq:Shift}.

Another possibility for future investigations would be to try to relate
the spectrum of fluctuation to certain cycles on the 
algebraic curve of a classical solution \cite{Beisert:2005di}. 
Similar results have been obtained in the context of matrix models 
and they might apply to this system in some form as well.

Finally, it could be interesting to study finite size effects 
at higher loops \cite{Beisert:2005fw} 
and for the proposed string Bethe equations \cite{Arutyunov:2004vx,Beisert:2004jw}.
These should be compared to native string theory computations
and would constitute a non-trivial test of the conjectured equations.%
\medskip%

\paragraph{Acknowledgements.}
We thank Volodya Kazakov, Charlotte Kristjansen, Matthias Staudacher and Arkady Tseytlin for discussions. 
N.~B.~would like to thank NORDITA for hospitality during work on this article.
The work of N.~B.~is supported in part by the U.S.~National Science
Foundation Grant No.~PHY02-43680. Any opinions,
findings and conclusions or recommendations expressed in this
material are those of the authors and do not necessarily reflect the
views of the National Science Foundation.
%N.B. would like to thank Nadav Drukker 
%for the friendly supply of coffee in times of need 
%and the cute little cups to have it in.

\bibliographystyle{nbshort}
\bibliography{Generic1J}

\begin{thebibliography}{10m}
\ifx\href\asklfhas\newcommand{\href}[2]{#2}\fi
\raggedright
\footnotesize
\parskip 0pt
\parindent -1em
\itemindent -1em
\itemsep 0pt

%%CITATION = HEP-TH 0212208;%%
\bibitem{Minahan:2002ve}
J.~A.~Minahan and K.~Zarembo,
\textsf{JHEP~0303,~013~(2003)},
\href{http://arXiv.org/abs/hep-th/0212208}{\texttt{hep-th/0212208}}.
%
%%CITATION = HEP-TH 0303060;%%
\bibitem{Beisert:2003tq}
N.~Beisert, C.~Kristjansen and M.~Staudacher,
\textsf{Nucl.~Phys.~B664,~131~(2003)},
\href{http://arXiv.org/abs/hep-th/0303060}{\texttt{hep-th/0303060}}.
%
%%CITATION = HEP-TH 0407277;%%
\bibitem{Beisert:2004ry}
N.~Beisert,
\textsf{Phys.~Rept.~405,~1~(2005)},
\href{http://arXiv.org/abs/hep-th/0407277}{\texttt{hep-th/0407277}}.
%
%%CITATION = HEP-TH 0307032;%%
\bibitem{Callan:2003xr}
C.~G.~Callan,~Jr., H.~K.~Lee, T.~McLoughlin, J.~H.~Schwarz, I.~Swanson and
  X.~Wu,
\textsf{Nucl.~Phys.~B673,~3~(2003)},
\href{http://arXiv.org/abs/hep-th/0307032}{\texttt{hep-th/0307032}}.
%
%%CITATION = HEP-TH 0505028;%%
\bibitem{Swanson:2005wz}
I.~Swanson,
\href{http://arXiv.org/abs/hep-th/0505028}{\texttt{hep-th/0505028}},
PhD thesis.
%
%%CITATION = HEP-TH 0204051;%%
\bibitem{Gubser:2002tv}
S.~S.~Gubser, I.~R.~Klebanov and A.~M.~Polyakov,
\textsf{Nucl.~Phys.~B636,~99~(2002)},
\href{http://arXiv.org/abs/hep-th/0204051}{\texttt{hep-th/0204051}}.
%
%%CITATION = HEP-TH 0204226;%%
\bibitem{Frolov:2002av}
S.~Frolov and A.~A.~Tseytlin,
\textsf{JHEP~0206,~007~(2002)},
\href{http://arXiv.org/abs/hep-th/0204226}{\texttt{hep-th/0204226}}.
%
%%CITATION = HEP-TH 0311139;%%
\bibitem{Tseytlin:2003ii}
A.~A.~Tseytlin,
\href{http://arXiv.org/abs/hep-th/0311139}{\texttt{hep-th/0311139}}.
%
%%CITATION = HEP-TH 0409147;%%
\bibitem{Beisert:2004yq}
N.~Beisert,
\textsf{Comptes~Rendus~Physique~5,~1039~(2004)},
\href{http://arXiv.org/abs/hep-th/0409147}{\texttt{hep-th/0409147}}.
%
%%CITATION = HEP-TH 0409296;%%
\bibitem{Tseytlin:2004xa}
A.~A.~Tseytlin,
\href{http://arXiv.org/abs/hep-th/0409296}{\texttt{hep-th/0409296}}.
%
%%CITATION = HEP-TH 0411191;%%
\bibitem{Zarembo:2004hp}
K.~Zarembo,
\textsf{Comptes~Rendus~Physique~5,~1081~(2004)},
\href{http://arXiv.org/abs/hep-th/0411191}{\texttt{hep-th/0411191}}.
%
%%CITATION = HEP-TH 0306130;%%
\bibitem{Frolov:2003tu}
S.~Frolov and A.~A.~Tseytlin,
\textsf{JHEP~0307,~016~(2003)},
\href{http://arXiv.org/abs/hep-th/0306130}{\texttt{hep-th/0306130}}.
%
%%CITATION = HEP-TH 0306139;%%
\bibitem{Beisert:2003xu}
N.~Beisert, J.~A.~Minahan, M.~Staudacher and K.~Zarembo,
\textsf{JHEP~0309,~010~(2003)},
\href{http://arXiv.org/abs/hep-th/0306139}{\texttt{hep-th/0306139}}.
%
%%CITATION = HEP-TH 0405167;%%
\bibitem{Freyhult:2004iq}
L.~Freyhult,
\textsf{JHEP~0406,~010~(2004)},
\href{http://arXiv.org/abs/hep-th/0405167}{\texttt{hep-th/0405167}}.
%
%%CITATION = HEP-TH 0405055;%%
\bibitem{Lubcke:2004dg}
M.~L{\"u}bcke and K.~Zarembo,
\textsf{JHEP~0405,~049~(2004)},
\href{http://arXiv.org/abs/hep-th/0405055}{\texttt{hep-th/0405055}}.
%
%%CITATION = HEP-TH 0408187;%%
\bibitem{Frolov:2004bh}
S.~A.~Frolov, I.~Y.~Park and A.~A.~Tseytlin,
\textsf{Phys.~Rev.~D71,~026006~(2005)},
\href{http://arXiv.org/abs/hep-th/0408187}{\texttt{hep-th/0408187}}.
%
%%CITATION = HEP-TH 0501203;%%
\bibitem{Park:2005ji}
I.~Y.~Park, A.~Tirziu and A.~A.~Tseytlin,
\textsf{JHEP~0503,~013~(2005)},
\href{http://arXiv.org/abs/hep-th/0501203}{\texttt{hep-th/0501203}}.
%
%%CITATION = HEP-TH 0502122;%%
\bibitem{Freyhult:2005fn}
L.~Freyhult and C.~Kristjansen,
\textsf{JHEP~0505,~043~(2005)},
\href{http://arXiv.org/abs/hep-th/0502122}{\texttt{hep-th/0502122}}.
%
%%CITATION = HEP-TH 0502173;%%
\bibitem{Beisert:2005mq}
N.~Beisert, A.~A.~Tseytlin and K.~Zarembo,
\textsf{Nucl.~Phys.~B715,~190~(2005)},
\href{http://arXiv.org/abs/hep-th/0502173}{\texttt{hep-th/0502173}}.
%
%%CITATION = HEP-TH 0502188;%%
\bibitem{Hernandez:2005nf}
R.~Hern\'andez, E.~L\'opez, A.~Peri\'a\~nez and G.~Sierra,
\href{http://arXiv.org/abs/hep-th/0502188}{\texttt{hep-th/0502188}}.
%
%%CITATION = PRLTA,74,816;%%
\bibitem{Sutherland:1995aa}
B.~Sutherland,
\textsf{Phys.~Rev.~Lett.~74,~816~(1995)}.
%
%%CITATION = HEP-TH 0503200;%%
\bibitem{Beisert:2005di}
N.~Beisert, V.~A.~Kazakov, K.~Sakai and K.~Zarembo,
\href{http://arXiv.org/abs/hep-th/0503200}{\texttt{hep-th/0503200}},
to appear in JHEP.
%
%%CITATION = HEP-TH 0402207;%%
\bibitem{Kazakov:2004qf}
V.~A.~Kazakov, A.~Marshakov, J.~A.~Minahan and K.~Zarembo,
\textsf{JHEP~0405,~024~(2004)},
\href{http://arXiv.org/abs/hep-th/0402207}{\texttt{hep-th/0402207}}.
%
%%CITATION = HEP-TH 0307042;%%
\bibitem{Beisert:2003yb}
N.~Beisert and M.~Staudacher,
\textsf{Nucl.~Phys.~B670,~439~(2003)},
\href{http://arXiv.org/abs/hep-th/0307042}{\texttt{hep-th/0307042}}.
%
%%CITATION = HEP-TH 0504190;%%
\bibitem{Beisert:2005fw}
N.~Beisert and M.~Staudacher,
\href{http://arXiv.org/abs/hep-th/0504190}{\texttt{hep-th/0504190}},
to appear in Nucl.~Phys.~B.
%
%%CITATION = HEP-TH 0406256;%%
\bibitem{Arutyunov:2004vx}
G.~Arutyunov, S.~Frolov and M.~Staudacher,
\textsf{JHEP~0410,~016~(2004)},
\href{http://arXiv.org/abs/hep-th/0406256}{\texttt{hep-th/0406256}}.
%
%%CITATION = HEP-TH 0409054;%%
\bibitem{Beisert:2004jw}
N.~Beisert,
\textsf{Fortschr.~Phys.~53,~852~(2005)},
\href{http://arXiv.org/abs/hep-th/0409054}{\texttt{hep-th/0409054}}.
%
\end{thebibliography}

\end{document}